\documentclass[11pt,a4paper,abstract=on]{scrartcl}

\usepackage{color}
\usepackage{ulem}
\usepackage{graphicx}

\usepackage{siunitx}
\usepackage{authblk}
\usepackage[breaklinks, colorlinks=true, citecolor=black, urlcolor=black, linkcolor=black]{hyperref}

\title{Radar Maxima: calibrated area-based probabilistic forecasts for heavy precipitation}

\author{Reinhold Hess}
\affil{Deutscher Wetterdienst, Offenbach, Germany}

\begin{document}

\maketitle

\begin{abstract}
We present, motivate, and evaluate Radar Maxima, a calibrated area-based probabilistic forecast product for heavy precipitation.
It is designed to overcome inherent limitations of point-based forecasts, which often yield low probabilities for extreme events due to spatial displacement errors.
The method aggregates radar-derived precipitation within 40\,km neighbourhoods
to statistically upscale forecasts from the ensemble system ICON-D2-EPS of DWD.

Evaluation considers both objective verification metrics and feedback from operational weather forecasters based on case studies.
Verification shows improved predictability, reliability and forecast sharpness.
Evaluation of forecasters confirmed operational value in some cases.

\end{abstract}

\section{Introduction}\label{sec:intro}

\begin{figure}[hbtp]
\centering
\includegraphics[width=0.49\textwidth]{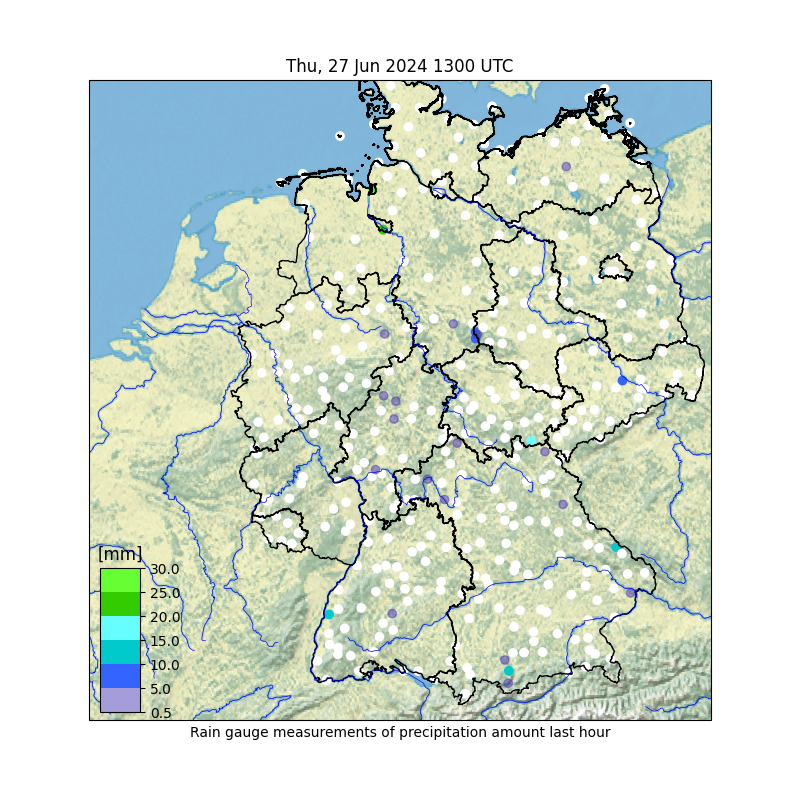}
\includegraphics[width=0.49\textwidth]{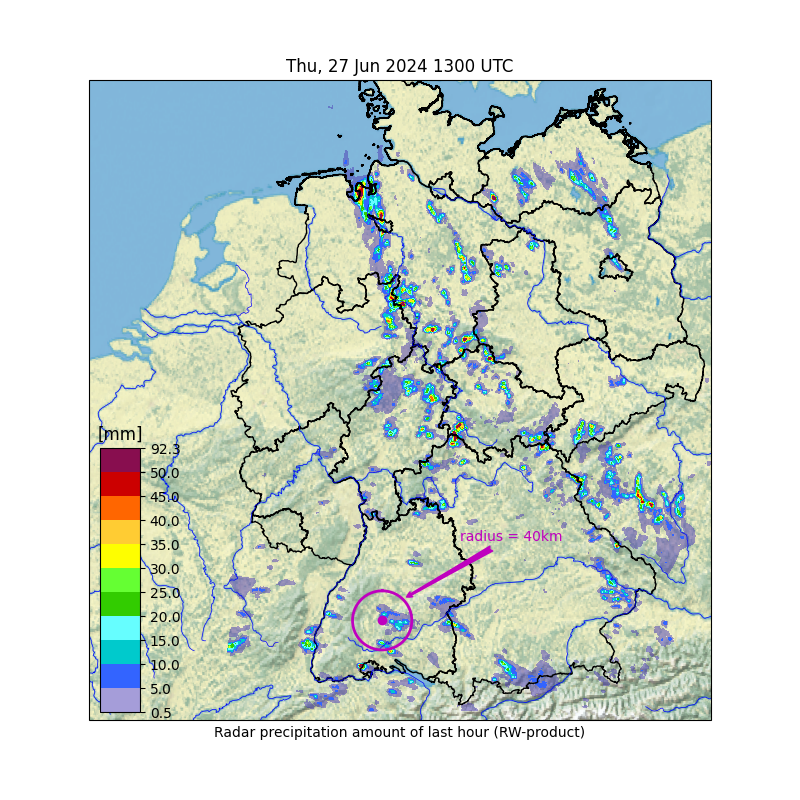}
\caption{
Left: synoptic rain gauge measurements for 27 June 2024, 13\,UTC.
Right: corresponding precipitation amounts from gauge-adjusted radar observations by DWD's operational radar network.
The circle with radius 40\,km around Station \texttt{Q651} (Hechingen) illustrates the spatial reference area used for Radar Maxima.
}\label{fig:radar}
\end{figure}

Heavy precipitation remains one of the most challenging variables to forecast in numerical weather prediction (NWP).
The difficulty arises from its inherent discrete and spatially discontinuous nature, especially under convective conditions.
In such scenarios, point measurements of rain rates provide limited information about surrounding regions,
as illustrated in Fig.~\ref{fig:radar}.
For instance, only about 17\% of one-hour rain rates exceeding 25\,mm are captured by synoptic observations in Germany \cite{lengfeld2020}.
Due to their spatial coverage, radar observations can reveal locations with much higher intensity and provide a better representation of extreme cases.
Average return periods of high-impact precipitation events detected via synoptic and radar observations
are summarised in Tab.~\ref{tab:freqocc}.
Events with more than 25\,mm/h are observed, on average, less than once every 4.34 years at an individual station,
while radar detects such events within a 40\,km radius approximately once every 4.25 months.

Current convection-resolving NWP models can simulate realistic precipitation patterns,
including heavy convective rainfall with reasonable precipitation amounts.
Nevertheless, due to initial errors and an exponential growth of forecast errors (i.e., the \emph{butterfly effect}),
predictability for the exact location (and timing) of convective cells remains limited, as the convective cells are often displaced.
In combination with scattered precipitation cells, these
displacements not only lead to missed events at the
correct location, but also to false alarms for areas where precipitation eventually does not occur (\emph{double penalty}).
Thus, straightforward grid-point-based evaluations of precipitation forecasts are inadequate.

Acknowledging this limitation of deterministic forecasts has motivated the development of ensemble forecasting,
where multiple model realisations are generated to account for initial uncertainties and model errors in
order to yield the full range of possible weather scenarios \cite{lewis2005}.
A set of 20 (ICON-D2-EPS) or even 50 (ECMWF-EPS) ensemble members raises the question of how to
effectively summarise the probabilistic information and communicate it to end  users \cite{fundel2019}\,.
Challenges include extracting useful information given the limited predictability,
focusing on the essential information for specific meteorological use cases and presenting probabilistic information in an intelligible way \cite{fundel2025}\,.
Various approaches have been implemented ranging from uni-variate grid-point based methods to the detection and processing of spatial and temporal objects.

The ensemble methods improve the situation for grid-point based evaluations,
as there is a better chance that at least a few ensemble members
predict the correct locations of heavy precipitation.
Nevertheless, the number of correct ensemble members will be most often small compared to the total number and
the resulting ensemble mean and estimated probabilities (i.e.\ the relative frequencies of correct ensemble members)
become small.

To mitigate these issues, spatial \emph{upscaling} techniques have been developed.
These approaches aggregate predictions over areas or use spatial combinations of nearby grid-points, commonly referred to as \emph{neighbourhoods}.
According to \cite{schwartz2017}, such approaches are categorised into \emph{Neighbourhood Ensemble Probability} (NEP) and
\emph{Neighbourhood Maximum Ensemble Probability} (NMEP) methods.
In NEP, neighbouring grid-points are treated as additional ensemble members for a given location
(\cite{theis2005}, \cite{roberts2008}, \cite{schwartz2010}),
thereby increasing the sample size for estimating probabilities via relative frequency.
However, the smoothing introduced by neighbourhood averaging reduces forecast sharpness and dampens probabilities for extreme events \cite{schwartz2017}.
In contrast, NMEP methods evaluate each ensemble member independently, determining whether an event (e.g., heavy precipitation)
occurs anywhere within a predefined neighbourhood.
The final probability is estimated from the relative frequency of ensemble members that detect the event within their neighbourhoods.
This approach avoids the loss of spatial correlation information inherent in grid-point-based methods
(see \cite{epstein1966}, \cite{krzysztofowicz1998}, \cite{kriesche2015}, \cite{kriesche2017b}).

An example of NMEP is provided by \emph{upscaled probabilities} \cite{benbuellegue2014}, visualised in Fig.~\ref{fig:upscale} (centre).
Here, maximum precipitation values within boxes of 10$\times$10 grid-points
(corresponding to 22$\times$22\,km$^2$ for the 2.2\,km resolution of ICON-D2)
are tested against a threshold (e.g., 10\,mm/h).
NMEP is particularly effective for rare, localised, and high-impact events for which point-scale probabilities are small \cite{roberts2023}.
A recent extension of NMEP for lightning forecasting is discussed in \cite{baumgartner2024}.

With the exception of \cite{baumgartner2024}, the aforementioned methods are based solely on raw ensemble forecasts and do not ensure that probabilistic outputs are statistically calibrated
(i.e., that forecast probabilities correspond to empirical frequencies).
The statistical connection between forecast probabilities and observed event frequencies improves reliability and trustworthiness of the forecasts
and supports rational, risk-based decision making \cite{wilks2001}.
Despite improvements with spatial upscaling techniques such as NEP and NMEP, the absence of calibration remains a critical limitation.
However, with calibration also \emph{sharpness} (i.e., the variability of the forecasts)
should be addressed in order to assess the value of a forecasting system in this context \cite{gneiting2007}.

\begin{figure}[hbtp]
\centering
\includegraphics[width=0.49\textwidth]{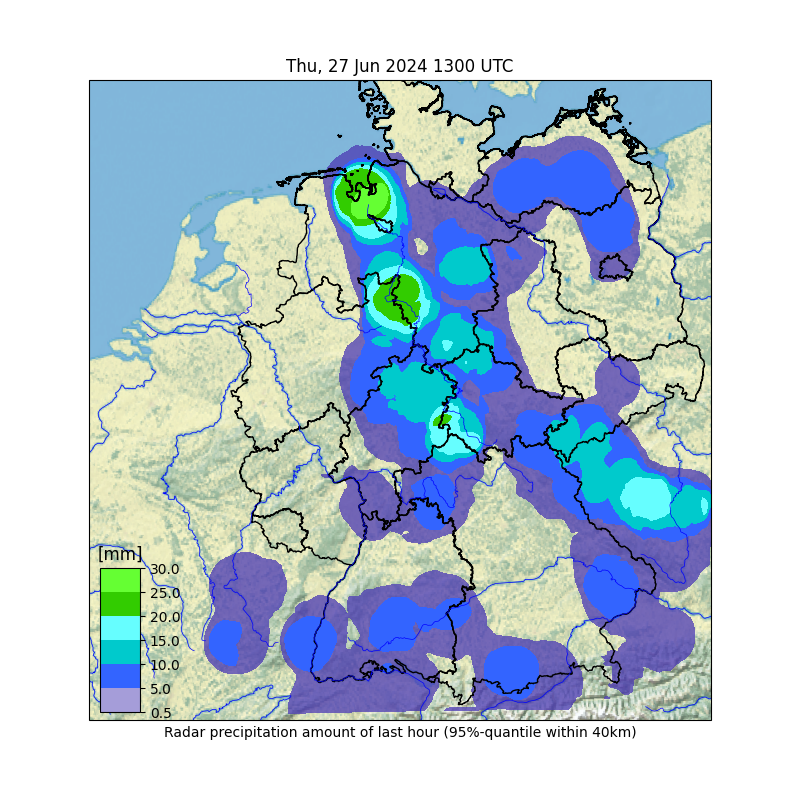}
\includegraphics[width=0.49\textwidth]{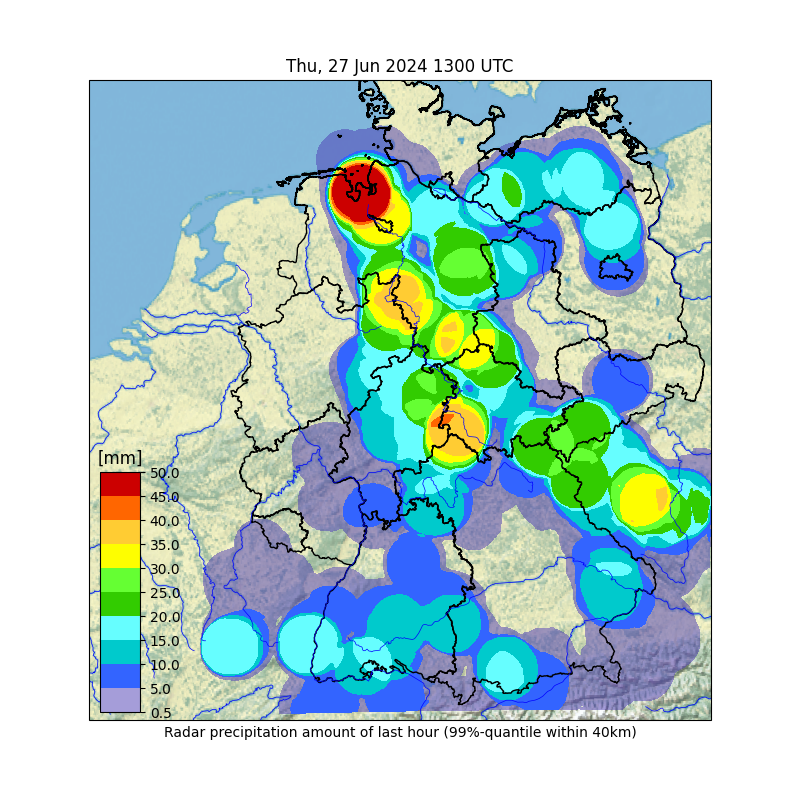}
\caption{
Gauge-adjusted radar precipitation (12-13\,UTC, 27 June 2024).
Left: 95\% spatial quantile within a 40\,km radius around each grid-point.
Right: corresponding 99\% spatial quantile.
}\label{fig:radarmax}
\end{figure}

This paper directly addresses the need of upscaling and calibration of probabilistic forecasts for heavy precipitation.
We propose \emph{Radar Maxima} as a forecast product based on statistical post-processing, which
uses historical radar-based spatial precipitation maxima as target variables and ensemble output from ICON-D2-EPS as predictors.
The precipitation maxima are derived in an NMEP approach as high quantiles (95\% and 99\%) of precipitation amounts
within a 40\,km radius (as shown in Fig.~\ref{fig:radarmax}) based on 
gauge-adjusted radar data (see Fig.~\ref{fig:radar}, right).

The training dataset for Radar Maxima is described in Sec.~\ref{sec:training_data}.
The statistical post-processing method is a Model Output Statistics (MOS) system,
which has been adapted to ensemble systems.%
\footnote{
The Radar Maxima development has been migrated recently to a neural network using dense layers and TensorFlow/PyTorch.
The results are similar, hence the findings in this paper using MOS are considered up to date.}
It is detailed in Sec.~\ref{sec:training}\@.
Evaluation is presented in Sec.~\ref{sec:evaluation}, including both a case study and statistical verification.
Following \cite{buizza2018}, we assess improvements in forecast accuracy and reliability,
comparing Radar Maxima to raw ensemble forecasts and to the upscaled probabilities mentioned above.
Finally, Sec.~\ref{sec:conclusion} provides a summary, including feedback from operational forecasters.

\section{Training Data}\label{sec:training_data}

Statistical modelling of Radar Maxima is carried out at the locations of 399 synoptic stations in Germany.
Although training with radar and model data could be performed at all model grid-points to produce gridded statistical forecasts,
restricting training to station locations reduces computational costs and avoids the need to interpolate statistical forecasts
for comparison and verification with conventional gauge measurements.

The predictands (targets) used for training are derived from radar data, as described in Sec.~\ref{sec:radar},
while the predictors (features) are based on operational ensemble forecasts from DWD, detailed in Sec.~\ref{sec:ensemble}\,.
The time series spans from 8 December 2010 to 31 December 2024;
however, only data up to 31 December 2023 are used for training, while data from 2024 are reserved exclusively for evaluation.

\subsection{Radar Data}\label{sec:radar}

The training targets are area-based precipitation amounts derived from DWD's operational radar network.
Precipitation estimates from reflectivity measurements are amount adjusted using data from approximately 1300 rain gauges in Germany \cite{winterrath2012}.
A composite of currently 17 operational radar stations provides hourly precipitation estimates on a 1$\times$1\,km$^2$ grid across Germany.
Spurious radar artefacts are removed using the cluster filter algorithm proposed in \cite{winterrath2007}.
\footnote{The cluster filter has been applied operationally since 31 August 2016 and manually to historical data before that.}
Apart from the filtering, the adjusted radar precipitation amounts agree with the rain gauge measurements at the locations of the synoptic stations.
An example of the filtered radar composites is shown in Fig.~\ref{fig:radar} (right).

At each training location, precipitation amounts within a 40\,km radius are considered;
the 95\% quantile of these values is taken as the maximum precipitation in the surrounding area.
This choice avoids the use of the absolute maximum (100\% quantile),
which can be distorted by residual erroneous radar pixels that remain after filtering.
The resulting radar maxima for all pixels are shown in Fig.~\ref{fig:radarmax} (left),
along with the corresponding 99\% quantiles for comparison (right).

To further characterise this derived radar product, Tab.~\ref{tab:freqocc} lists average return periods
for events where hourly precipitation exceeds various thresholds.
Compared to synoptic rain measurements, spatial 95\% quantiles occur more frequently at lower thresholds (up to 25\,mm/h).
At higher thresholds, however, return periods become longer than those of synoptic observations.
This is because extreme rainfall events are often highly localised, rarely affecting more than 5\% of the 40\,km circles.
This behaviour is clearly seen when comparing with the 99\% quantiles, which show much shorter return periods.

\begin{table}[hbtp]
\caption{Average return periods per station for precipitation events exceeding various thresholds,
based on conventional synoptic rain gauges and on the 95\% and 99\% spatial quantiles within 40\,km circles
derived from gauge-adjusted radar estimates.
The statistics cover 14 years (8 Dec 2010 to 31 Dec 2024) and 399 stations.
Multiple events at the same station on the same day are counted only once.
Note: each 40\,km circle contains, on average, 5.1 stations (including the central station),
so the same event may be counted multiple times if detected at different stations.
SYNOP: precipitation from synoptic rain gauges;
RADAR~(95\%) and RADAR~(99\%): 95\% and 99\% quantiles of radar precipitation amounts within a 40\,km radius.
}\label{tab:freqocc}
\begin{center}
\begin{tabular}{rc*{2}{l}}
Threshold & \hspace{2ex}SYNOP & RADAR (95\%) & RADAR (99\%)\\
\hline
5\,mm/h  & 1.09\,months & 11.1\,days    & 5.89\,days \\
10\,mm/h & 4.35\,months & 1.61\,months  & 16.7\,days \\
15\,mm/h & 11.4\,months & 5.16\,months  & 1.15\,months \\
20\,mm/h & 2.11\,years  & 1.34\,years   & 2.21\,months \\
25\,mm/h & 4.34\,years  & 4.27\,years   & 4.25\,months \\
30\,mm/h & 8.48\,years  & 10.8\,years   & 8.32\,months \\
35\,mm/h & 16.5\,years  & 25.6\,years   & 1.35\,years \\
40\,mm/h & 30.0\,years  & 41.3\,years   & 2.52\,years \\
\hline
\end{tabular}
\end{center}
\end{table}

\subsection{Ensemble Data}\label{sec:ensemble}

For training and evaluation, since 8 December 2010 more than 14 years of ensemble forecasts have been collected,
including data from COSMO-DE-EPS, COSMO-D2-EPS, and ICON-D2-EPS throughout their operational life-cycles.
These convection-permitting ensemble models have resolutions of 2.8$\times$2.8\,km$^2$ (COSMO-DE-EPS)
and 2.2$\times$2.2\,km$^2$ (COSMO-D2-EPS and ICON-D2-EPS).
An overview of the model timelines is provided in \cite{lerch2025}, and
technical details can be found in \cite{baldauf2011}, \cite{gebhardt2011}, \cite{peralta2012}, and \cite{reinert2023}\,.
Model changes and changes in model configurations are accepted to ensure long time series,
which are essential for modelling and evaluating rare and extreme precipitation events.
Tab.~\ref{tab:freqocc} demonstrates empirical return periods for hourly precipitation events up to 40\,mm,
which occur statistically only once every 30 years per station.

The ensemble forecasts are used as input for statistical training and are mapped to the locations of the synoptic stations
using the nearest-neighbour grid-point method, avoiding any smoothing.
To account for the high spatial variability of precipitation,
mean and standard deviation of two spatial variants are used additionally:
medium scale (MS, 6$\times$6 grid-points) and large scale (LS, 11$\times$11 grid-points),
centred around each station.
The dataset includes 55 model variables (145 variables when including MS and LS variants) for all 20 ensemble members.
A detailed description of the dataset (subset for 170 stations) is provided in \cite{lerch2025}\,.

\section{Statistical training with MOS}\label{sec:training}

Statistical training is performed using a MOS system
based on \cite{knuepffer1996}, which has been extended for ensemble data and probabilistic forecasts, including logistic regression \cite{hess2020}.

Most of the predictors are
ensemble mean, standard deviation, and the 75\% and 90\% ensemble quantiles of the data set described in Sec.~\ref{sec:ensemble}, but
additional variables are derived as well, such as layer thicknesses, CAPE, or SWEAT indices.
Moreover, the most recent available synoptic and radar observations (in the form of the 40\,km spatial quantiles) are defined as persistence predictors.
This is meaningful in an operational context, as the output of ICON-D2-EPS becomes available only about 2\,h after the nominal model start,
at which point new observations are already available.
Additionally, when modelling the time steps one after another, the MOS forecasts of the preceding steps
can be used as predictors for the current time step, which also helps the statistical model to incorporate meteorological persistence
and to stabilise the forecasts in time.

Precipitation probabilities are trained in two steps:
first, precipitation amounts are estimated for each station, forecast run, and time step individually using linear regression;
second, the probabilities that these amounts exceed predefined thresholds are modelled thereafter using global logistic regression for all stations simultaneously.

\begin{figure}[hbtp]
\center
\includegraphics[width=0.4\textwidth]{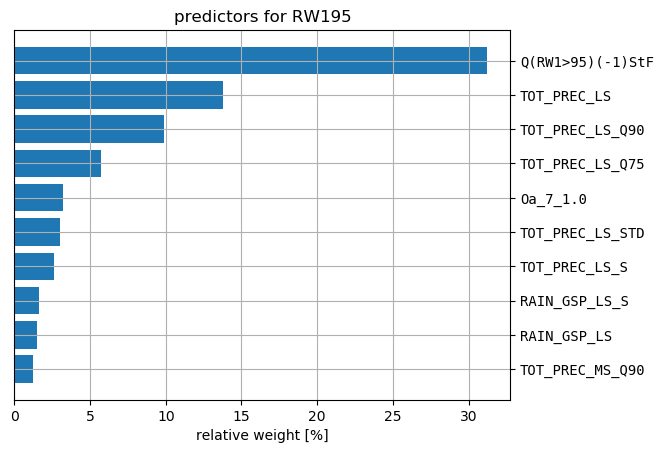}
\tiny
\begin{tabular}{ll}
\textbf{Predictor name} & \textbf{Short description}\\
\texttt{Q(RW1>95)(-1)StF} & Statistical forecast of 95\%~quantile for the previous hour\\
\texttt{TOT\_PREC\_LS} & Model total precipitation, spatial mean (11$\times$11 grid-points)\\
\texttt{TOT\_PREC\_LS\_Q90} & Model total precipitation, 90\% ensemble quantile of spatial means (11$\times$11 grid-points)\\
\texttt{TOT\_PREC\_LS\_Q75} & Model total precipitation, 75\% ensemble quantile of spatial means (11$\times$11 grid-points)\\
\texttt{Oa\_7\_1.0} & Most recent available radar-derived maxima from observations\\
\texttt{TOT\_PREC\_LS\_STD} & Model total precipitation, ensemble standard deviation of spatial means (11$\times$11 grid-points)\\
\texttt{TOT\_PREC\_LS\_S} & Model total precipitation, ensemble mean of spatial standard deviations (11$\times$11 grid-points)\\
\texttt{RAIN\_GSP\_LS\_S} & Model large-scale rain, ensemble mean of spatial standard deviations (11$\times$11 grid-points)\\
\texttt{RAIN\_GSP\_LS} & Model large-scale rain, spatial mean (11$\times$11 grid-points)\\
\texttt{TOT\_PREC\_MS\_Q90} & Model total precipitation, 90\% ensemble quantile of spatial means (6$\times$6 grid-points)\\
\end{tabular}
\hfill
\caption{Most relevant predictors of the MOS system for Radar Maxima according to their relative weights.
Weights are aggregated across 399 statistical equations (one per station, forecast run, and lead time)
during the training period from 8 December 2010 to 31 December 2023.
While the selected predictors appear highly correlated and redundant, this results from the compilation of many MOS equations for this statistics.
Within each individual equation, predictors are selected based on significance testing and strongly correlated predictors exclude each other.}
\label{fig:predictors}
\end{figure}

During the first training step
the most significant predictors of the precipitation amounts are selected from a pool of approximately 400 potential predictors
int total using Students $t$-tests.
Figure~\ref{fig:predictors} displays the most relevant predictors for the spatial 95\%~quantile
of the 00\,UTC forecast run, aggregated over all forecast steps, according to their relative weights in the MOS equations.
The variable names are technical and refer to specific configurations.
The most important predictor, \texttt{Q(RW1>95)(-1)StF}, represents the statistical forecast of the 95\%~quantile at the previous time step.
The following predictors are various derivatives of the model's total precipitation (\texttt{TOT\_PREC}),
while \texttt{Oa\_7\_1.0} corresponds to the most recent radar observation at 02\,UTC.
This predictor is only relevant for the first few time steps;
thereafter, the forecasts from preceding steps gain more impact.
Overall, most selected predictors are derived from model precipitation, as expected;
however, additional variables such as the \texttt{SWEAT} and \texttt{SMS} indices and also relative humidity
contribute to the regression and are used at individual stations.

The second training step applies logistic regression for the precipitation probabilities, using only the modelled quantiles from the first step as predictors.
The predictands for the logistic regression are binary and defined as 1 if the spatial quantiles based on radar observations exceed the threshold, and 0 otherwise.
For thresholds up to 3\,mm/h, this is again performed for each station and time step individually;
for thresholds of 5\,mm/h and above, statistical sampling is improved by modelling all stations jointly.

\section{Evaluation}\label{sec:evaluation}

The Radar Maxima product was evaluated by the ESSL Testbed in 2023 \cite{pucik2023} and 2024 \cite{pucik2024}
from the perspective of operational forecasters.
During four weeks in June and July each of both years, Radar Maxima and other short-term forecast products were
assessed based on the present weather conditions, with performance reviewed on the following day.
This process yielded several case studies involving extreme precipitation and other significant weather.
One such case study from 2024 is presented in Sec.~\ref{sec:cases}.
However, as case studies primarily illustrate individual events, mostly involving severe weather,
verification statistics for the entire year 2024 are provided in Sec.~\ref{sec:verification}.
This includes non-extreme weather situations as well and provides a more comprehensive assessment of Radar Maxima performance.

\subsection{Case study}\label{sec:cases}

\begin{figure}[hbtp]
\centering
\includegraphics[width=0.49\textwidth]{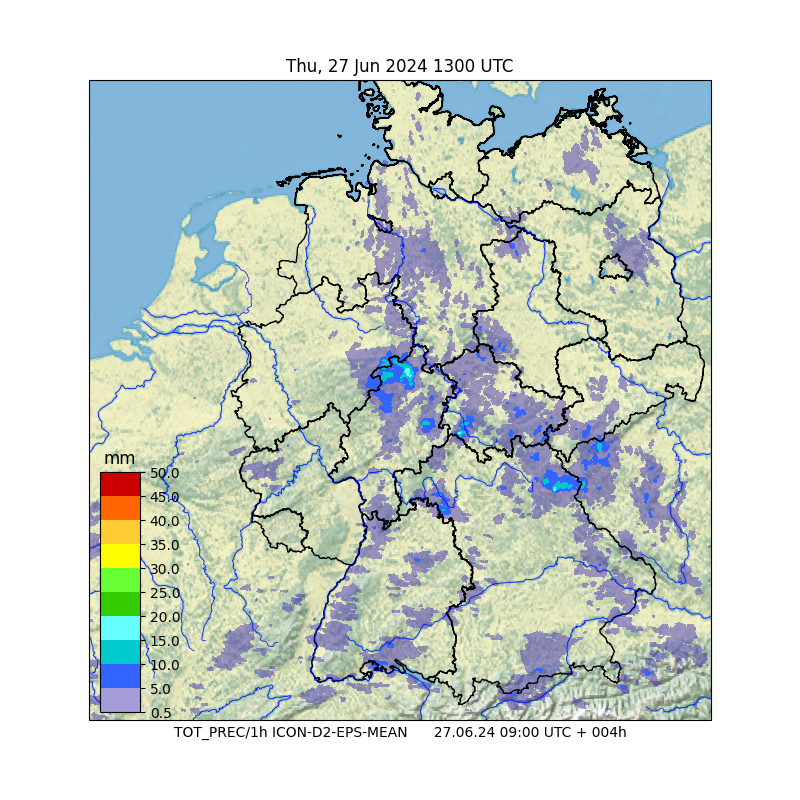}
\includegraphics[width=0.49\textwidth]{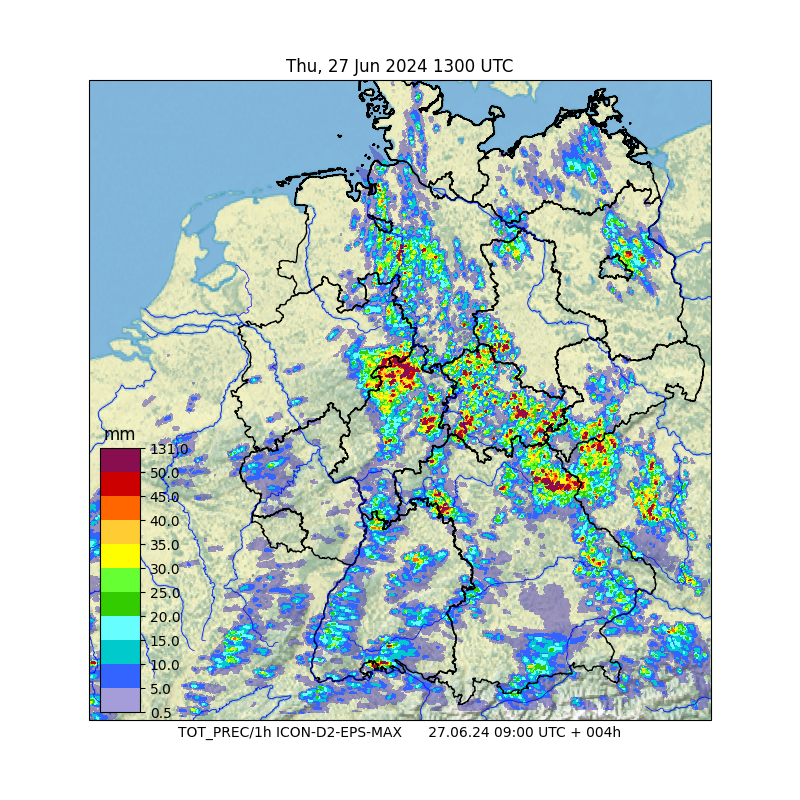}
\caption{
ICON-D2-EPS forecasts of total precipitation on 27 June 2024 between 12\,UTC and 13\,UTC based on grid-points.
Left: ensemble mean; right: ensemble maximum.}\label{fig:id2eps}
\end{figure}

For conciseness, we focus on one-hour accumulated precipitation from 12\,UTC to 13\,UTC on
27 June 2024, as shown in Figs.~\ref{fig:radar} and \ref{fig:radarmax}\,.
We compare 4\,h forecasts from Radar Maxima to the raw (untrained) ICON-D2-EPS forecasts started at 09\,UTC,
which serves as the input for Radar Maxima in this case.
The pointwise ensemble mean and maximum from ICON-D2-EPS are shown in Fig.~\ref{fig:id2eps}\,.

\begin{figure}[hbtp]
\centering
\includegraphics[width=0.3\textwidth]{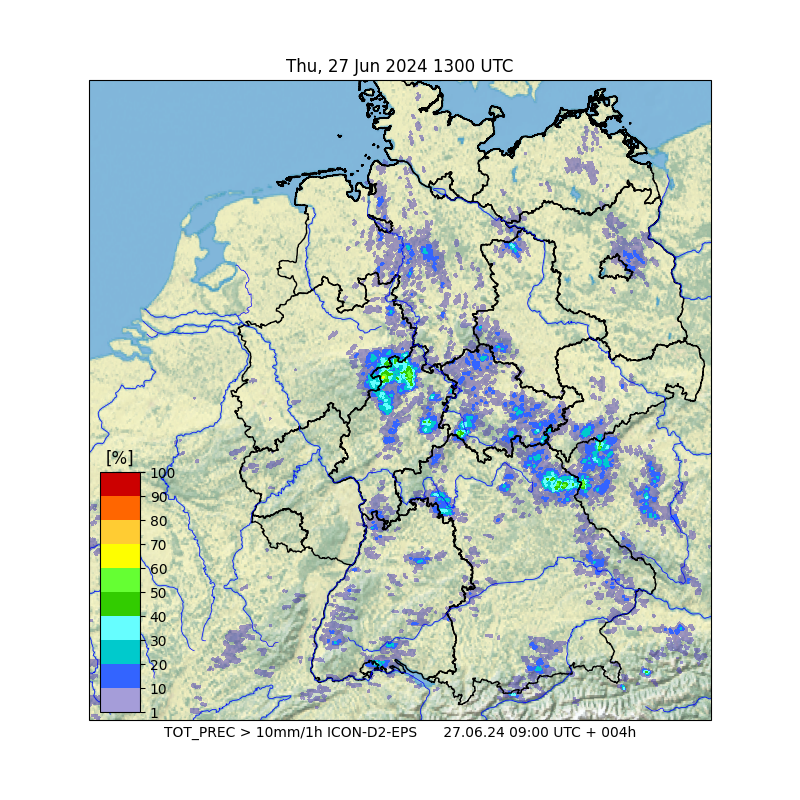}
\includegraphics[width=0.3\textwidth]{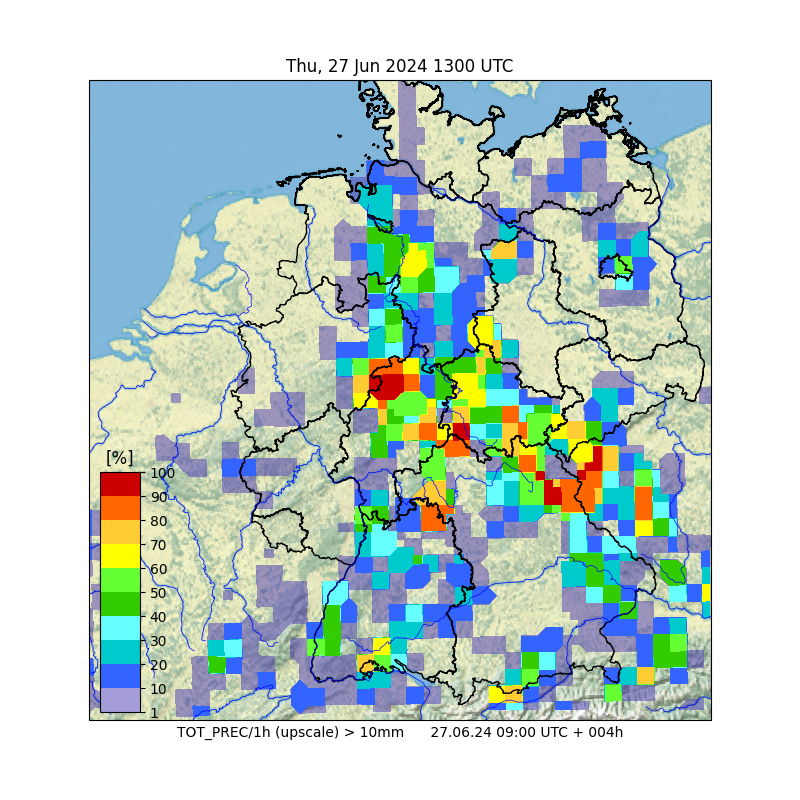}
\includegraphics[width=0.3\textwidth]{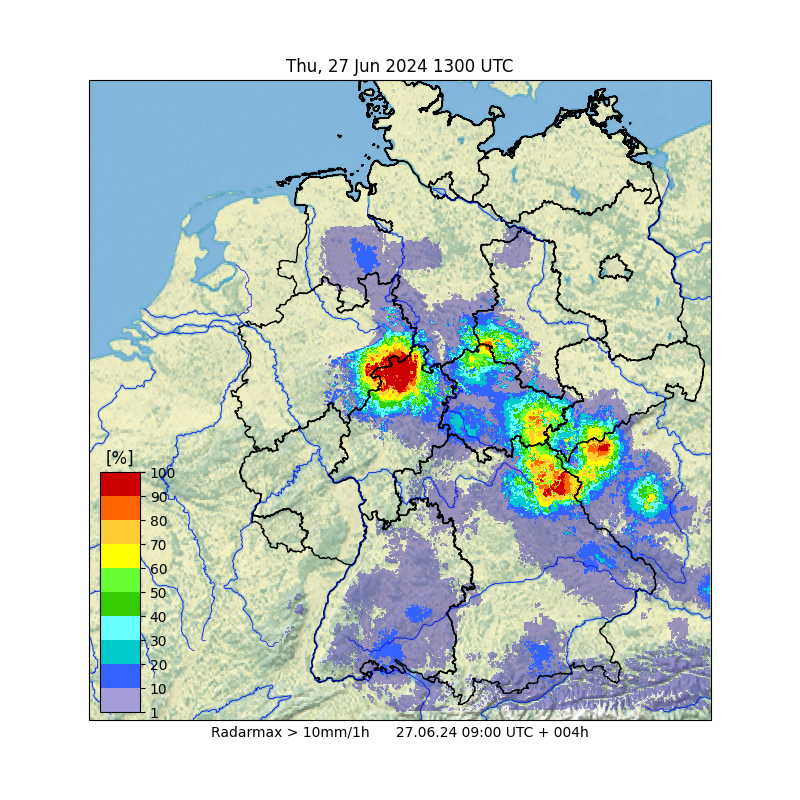}
\caption{
Estimated probabilities of exceeding 10\,mm/h precipitation on 27 June 2024 between 12\,UTC and 13\,UTC
based on ICON-D2-EPS started at 09\,UTC.
Left: pointwise relative frequencies of ensemble members with total precipitation (TOT\_PREC) above the threshold;
centre: corresponding upscaled probabilities over 22$\times$22\,km$^2$ boxes;
right: corresponding Radar Maxima estimate.}\label{fig:upscale}
\end{figure}

The ensemble mean forecast offers damped precipitation values, peaking at 18.2\,mm,
which fail to represent the observed maximum rain rates of up to 92.3\,mm/h (Fig.~\ref{fig:radar}).
This underestimation results from the non-Gaussian distribution of precipitation
with generally high probabilities for small precipitation amounts, as discussed in Sec.~\ref{sec:intro}.
Conversely, the pointwise ensemble maximum overestimates the precipitation, with rates up to 131\,mm/h,
and displays widespread signals not corroborated by radar estimates.
The pointwise probability that precipitation exceeds 10\,mm/h is
computed as the relative frequency of ensemble members exceeding the threshold at each grid-point,
shown in Fig.~\ref{fig:upscale} (left). The maximum probability is 65\%, an underwhelming signal
when compared to radar estimates, which reveal many locations exceeding 20\,mm/h.

A useful benchmark for Radar Maxima with spatial reference are \emph{upscaled probabilities}
for precipitation \cite{benbuellegue2014}.
Here, the maximum precipitation within boxes of
22$\times$22\,km$^2$ (corresponding to 10$\times$10 grid-points)
is evaluated for each ensemble member.
The probability that precipitation exceeds 10\,mm/h is estimated as the relative frequency
of ensemble members whose grid-box maxima exceed this threshold.
This benchmark product is shown in Fig.~\ref{fig:upscale} (centre).
The spatial reference of 22$\times$22\,km$^2$ accounts for
the high spatial variability of precipitation.
These upscaled probabilities yield stronger and more informative signals,
however, they are less selective: many regions are indicated with high probabilities
where no precipitation is eventually observed. This uncalibrated product does not take model errors into account.

Radar Maxima, on the other hand, also provides strong signals up to 100\% for high precipitation rates exceeding 10\,mm/h,
but in a much more selective manner.
The statistical product suppresses signals from ICON-D2-EPS in situations
that are assessed as questionable due to model errors identified during statistical training.
Despite this, some spatial displacements remain and
the most intense precipitation at the north German coast is completely missed.
This reflects limitations of the underlying NWP model,
as statistical post-processing cannot compensate for missing signals in the input.

\subsection{Verification}\label{sec:verification}

As long as probabilistic forecasts are neither 0\% nor 100\%, they
cannot be evaluated strictly as correct or incorrect.
If an event with low forecast probability occurs, this does not necessarily imply the forecast system is poor;
it simply indicates that it is \emph{unlikely} that the forecast system is good.
If the event does not occur in these cases, the forecast system may appear better,
but this could happen only by chance.
The inverse is true if the forecast shows high probabilities.
Assessing probabilistic forecasts requires evaluation over many cases
to allow for a statistically significant assessment of forecast system quality, therefore.

Radar Maxima was trained using data from December 2010 through December 2023.
Forecasts for 2024 are used for verification and are compared with raw (untrained) ICON-D2-EPS forecasts.
Rare events are inherently challenging for both training and verification due to their low frequency.
Even though statistics are aggregated over an entire year,
verification is limited for the 5\,mm/h threshold;
for higher thresholds, statistical significance decreases, notice the large return periods for heavy precipitation in Tab.~\ref{tab:freqocc}.

Forecast accuracy and predictability are analysed in Section~\ref{sec:acc} using the Brier Score (BS)
and its corresponding skill score. Calibration, reliability, and sharpness are addressed in Section~\ref{sec:calibration}.

\subsubsection{Accuracy and Predictability}\label{sec:acc}

Let $f_i$, $i\!=\!1,\ldots,n$, denote the forecast probabilities for exceeding a given precipitation threshold,
with sample size $n$.
The corresponding observations $o_i$ are defined as 1 if the event occurred and 0 otherwise. The BS is then defined as:
\begin{equation}\label{eq:brier}
\mathit{BS} = \frac{1}{n}\sum_{i=1}^{n} (f_i - o_i)^2\,.
\end{equation}

The BS basically represents the mean squared error of the forecast probabilities.
Since forecasting difficulty varies with weather regimes,
the BS score is typically normalised using a reference forecast, yielding the Brier Skill Score (BSS).
A straightforward reference forecast is the climatological frequency of the selected time period,
i.e. the base rate of the verification sample
$\bar{o} = \frac{1}{n}\sum_{i=1}^{n}o_i$.
Applying a constant forecast $f_i = \bar{o}$ yields the reference BS$_{ref}$:
\begin{equation}\label{eq:brierref}
\mathit{BS_{ref}} = \frac{1}{n}\sum_{i=1}^{n} (\bar{o} - o_i)^2 = \bar{o}(1 - \bar{o})\,.
\end{equation}
Note: forecasting the base rate is not as trivial as it appears.
The term $\bar{o}(1\! -\! \bar{o})$, also known as \emph{uncertainty}, reflects the inherent variability of the observations.
The BSS is defined as:
\begin{equation}\label{eq:bss}
\mathit{BSS} = 1 - \frac{\mathit{BS}}{\mathit{BS_{ref}}}\,,
\end{equation}
which equals 0 when the forecast matches the reference and 1 when it is perfect.
The BSS typically decreases with lead time and
predictability has finished, when BSS approaches zero.

\begin{figure}[hbtp]
\centering
\includegraphics[width=0.4\textwidth]{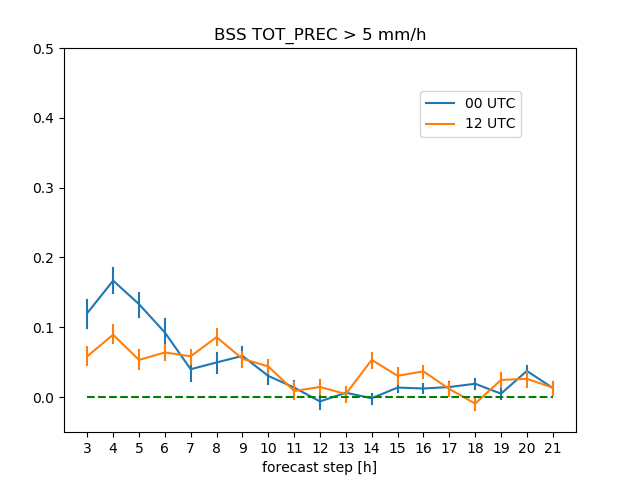}
\includegraphics[width=0.4\textwidth]{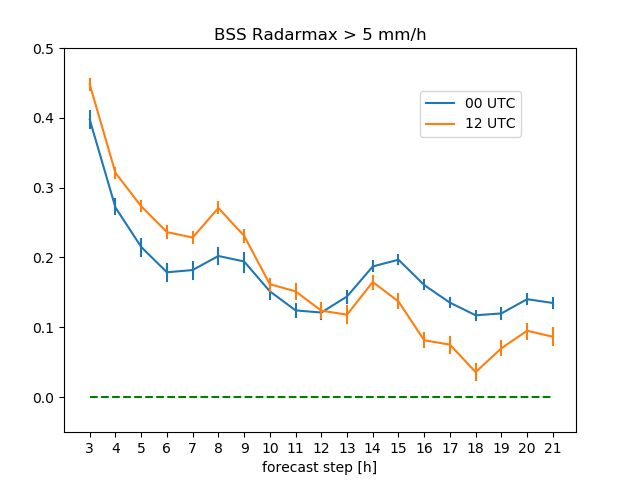}
\caption{Brier Skill Score for raw model total precipitation (TOT\_PREC) $>$ 5\,mm/h (left) and Radar Maxima $>$ 5\,mm/h (right) as a function of forecast lead time from 00 and 12\,UTC. Vertical bars show the interquartile range (25\%--75\%) from bootstrapping. Dashed line at zero marks the limit of predictability relative to their own sample climatology. Statistics are based on 399 stations with forecasts from 2024.}\label{fig:stepstat}
\end{figure}

Figure~\ref{fig:stepstat} shows BSS for grid-point precipitation forecasts from ICON-D2-EPS
and from Radar Maxima at the 5\,mm/h threshold,
evaluated as a function of lead time.
Reference forecasts are sample base rates in both cases. ICON-D2-EPS is verified using pointwise observations,
while Radar Maxima uses the spatial 95\% quantiles within a 40\,km radius equivalent to
its training targets.

The raw model precipitation shows limited skill beyond 12 hours,
while Radar Maxima maintains significant skill up to the maximum forecast horizon (21 hours).
This improvement is due to both statistical calibration and spatial aggregation, though their individual contributions are difficult to isolate.

\subsubsection{Calibration and Sharpness}\label{sec:calibration}

\begin{figure}[hbtp]
\centering
\includegraphics[width=0.3\textwidth]{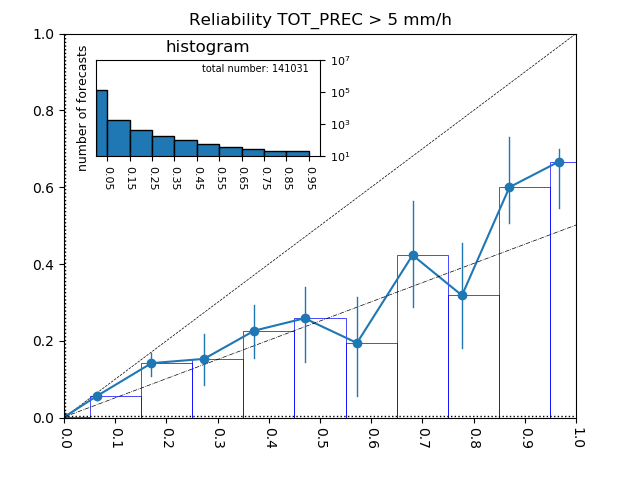}
\includegraphics[width=0.3\textwidth]{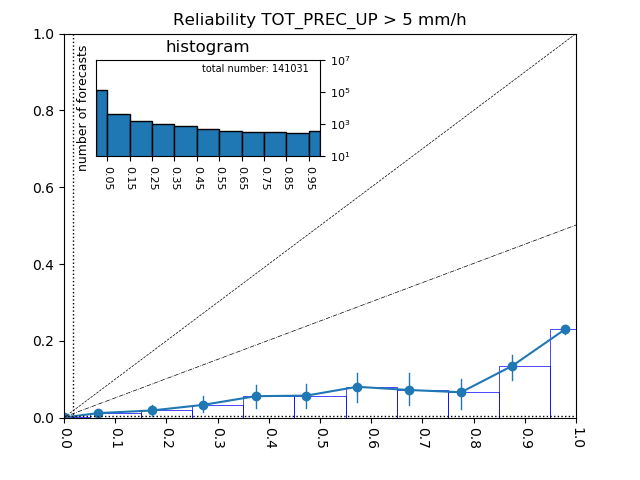}
\includegraphics[width=0.3\textwidth]{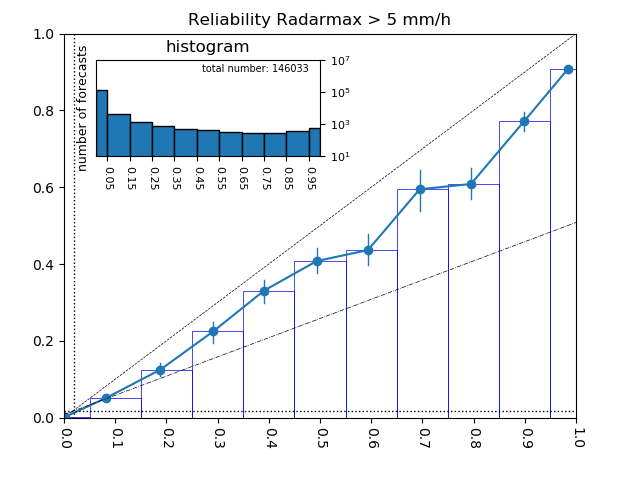}
\caption{Observed relative frequencies vs. forecast probabilities for 3-hour forecasts
starting at 12\,UTC for a 5\,mm/h threshold.
Left: raw grid-point model precipitation; centre: upscaled model probabilities;
right: Radar Maxima.
Vertical bars indicate 5\% to 95\% confidence intervals via bootstrapping.
Vertical and horizontal dotted lines show base rates of forecast and observed probabilities.
The \emph{no skill line} is dotted and inclined; relative frequencies above it
indicate a positive Brier Skill Score.
Grid point and upscaled precipitation both are verified against synoptic gauge measurements, while
Radar Maxima are compared to radar derived 95\% spatial quantiles within a 40\,km radius.
Statistics are based on 399 stations with forecasts for 2024.}\label{fig:reliability}
\end{figure}

Statistical calibration refers to the consistency between the predicted probability distributions and observed outcomes \cite{gneiting2007}.
A well-calibrated probabilistic forecast is bias free compared to observed relative frequencies.
This can be assessed using reliability diagrams as shown in Fig.~\ref{fig:reliability}
for raw grid-point based and upscaled ICON-D2-EPS forecasts, as well as for Radar Maxima.
ICON-D2-EPS grid-point based and upscaled forecasts are verified against pointwise gauge data; Radar Maxima is evaluated against the radar-derived 95\% quantiles as used in training.

The reliability diagram for raw grid-point precipitation (left) reveals pronounced over-forecasting
(given the forecast probability is larger than 5\%)
and almost no significant performance above the \emph{no-skill line}.
The upscaled precipitation product drastically over-predict the grid-point observations,
which is explainable due to the use of pointwise verification data.
Thus, the upscaled forecasts cannot be used to estimate grid-point model precipitation.
In contrast, Radar Maxima (right) still over-forecasts
(for probabilities larger than 5\%)
but demonstrates improved reliability and a more balanced forecast probability histogram.
It should be noted that there is no unconditional over-forecasting neither for
grid-point precipitation nor for Radar Maxima, as overall base rates
$\bar{f} = \frac{1}{n}\sum_{i=1}^{n}f_i$ and $\bar{o}$ 
almost coincide in both cases.
(The large fraction of forecasts with very small probabilities
underestimates verification data and counterbalances the overall statistics.)

The objective of probabilistic forecasting is to maximise \emph{sharpness} (i.e., the variability of the forecasts)
subject to calibration \cite{gneiting2007}.
While the base rate of the selected time period is perfectly calibrated,
it lacks informativeness and is useful only as reference.
Ideally, forecast sharpness should match observation sharpness, which implies that
forecast probabilities are either 0 or 1 in the same frequency as the observations
(not necessarily identical).
Here, sharpness is measured by the standard deviation of forecast probabilities
$\sigma_f = \frac{1}{n}\sqrt{\sum_{i=1}^n\!{\left(f_i\!-\!\bar{f}\right)^2}}$ to
be consistent with the standard deviation of observations, which is
$\sigma_o=\sqrt{\mbox{uncertainty}}=\sqrt{\bar{o}(1\!-\!\bar{o})}$.

\begin{figure}[hbtp]
\centering
\includegraphics[width=0.4\textwidth]{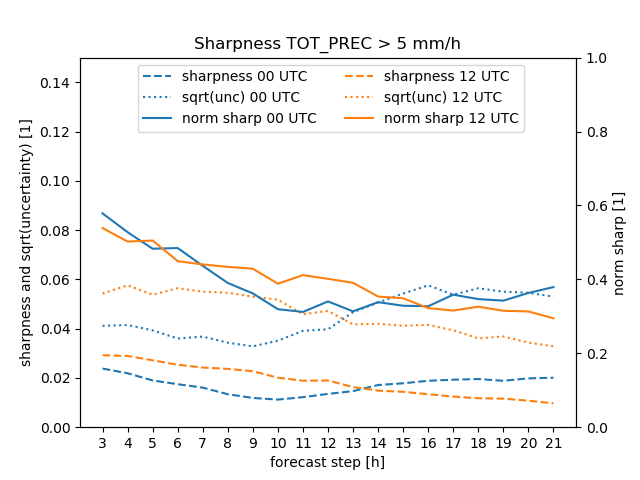}
\includegraphics[width=0.4\textwidth]{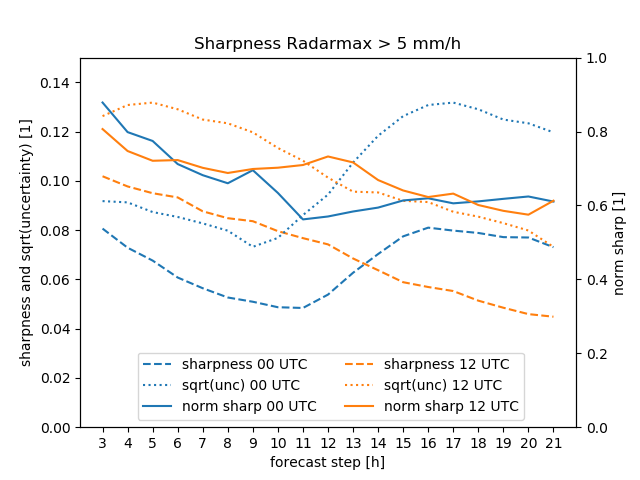}
\caption{Forecast sharpness ($\sigma_f$), observation variability ($\sigma_o=\sqrt{\mbox{uncertainty}}$),
and normalised sharpness ($\sigma_f / \sigma_o$) for a 5\,mm/h threshold.
Left: grid-point precipitation; right: Radar Maxima.
Forecasts initialised at 00 and 12\,UTC.
Statistics are based on 399 stations with forecasts for 2024.}\label{fig:sharpness}
\end{figure}

Figure~\ref{fig:sharpness} shows forecast and observation sharpness and
normalised forecast sharpness ($\sigma_f / \sigma_o$)
for grid-point precipitation and Radar Maxima forecasts at the 5\,mm/h threshold.
For 12\,UTC forecasts, sharpness declines with lead time in both cases.
For 00\,UTC runs, it decreases until about 10--11\,UTC and increases afterwards,
reflecting the diurnal cycle of convective precipitation.
Normalised sharpness compensates for the daily cycle of precipitation and reveals 
how much forecast variability and potential information content is lost compared to a perfect forecast (i.e., the observations).

Radar Maxima consistently demonstrates higher sharpness than the raw grid-point forecasts of ICON-D2-EPS.
Also uncertainty (and its square root) is higher for Radar Maxima due to its spatial upscaling
of precipitation.
Its normalised sharpness declines from 80\% to 60\% over time,
while that of the raw model falls from below 60\% to 30--40\% demonstrating
a smaller loss of forecast variability for the upscaled and calibrated product.

\section{Conclusions}\label{sec:conclusion}

Radar Maxima was introduced as an area-based forecast product designed to overcome the inherent limitations
of point-based precipitation forecasts,
in particular spatial displacements of precipitation and the restricted predictability and reliability in probabilistic forecasts.
The product aggregates probabilistic rainfall information from DWD's numerical ensemble,
provides a rapid overview of the current weather situation,
and is intended to support operational weather forecasters in identifying
critical areas worthy of detailed inspection.
Calibration with observations enables statistical correction for ensemble model errors,
providing an objective basis for weather warnings and supporting risk-based decision-making.
Over-forecasting still occurs whenever a chance for precipitation is detected (when forecast probabilities are larger than 5\%), nevertheless,
compared to uncalibrated grid-point forecasts, predictability, reliability, and sharpness are significantly improved.

Evaluation within the ESSL-Testbed \cite{pucik2023,pucik2024} confirmed
that Radar Maxima has value and is useful in specific situations%
\footnote{Problems with artefacts in areas of complex topography, reported in 2023, were solved in 2024.}.
The sizes of the highlighted precipitation areas were generally considered acceptable.
However, some criticism concerned the tendency of the calibration to reduce probabilities excessively compared to uncalibrated products,
such as maximum pointwise model precipitation or the upscaled forecasts introduced in Sec.~\ref{sec:intro}.
In such cases, the added value over the uncalibrated products was not always evident.

These reviews are understandable given forecaster's focus on extreme events
 and their sensitivity to missed alarms.
Strong forecast signals are expected when heavy precipitation occurs.
Considerable over-forecasting is often tolerated, as are spatial and temporal displacements of precipitation forecasts.
(For example, there were cases in which the raw upscaled probabilities predicted precipitation with 100\% certainty,
yet no precipitation was observed; in such cases, Radar Maxima meaningfully reduced the probabilities according to model errors represented in the training data.)
This evaluation bias is well-documented as the \emph{Forecaster's Dilemma} \cite{lerch2015},
which highlights that restricting evaluation to extreme observations is insufficient
for assessing forecast skill.
Ultimately, forecast accuracy and skill not only depend on the probability of detection, but also on the false alarm ratio and the extent of over-forecasting.

Radar Maxima missed some heavy precipitation events, for example at the north coast of Germany.
This shortcoming was due to a missing signal in the ICON-D2-EPS input data
(compare Figs.~\ref{fig:radar} and \ref{fig:upscale}).
By contrast, ICON-RUC%
\footnote{ICON-RUC is a rapid-update-cycle version of ICON with improved micro-physics, developed within the SINFONY project at DWD \cite{DWD_SINFONY}.}
predicted substantial precipitation in this case (not shown).
As a post-processing system, Radar Maxima cannot compensate for missing input signals,
and would benefit from improved numerical model input.
However, it remains an open question how much data from a new model should be available for retraining.

The neighbourhood radius of 40\,km and the spatial quantile of 95\% are tunable parameters of Radar Maxima.
While the spatial scale appears to match forecaster's expectations,
the 95\% quantile may be too low.
Especially extreme precipitation cells often have small spatial extents covering less than 5\% of the 40\,km radius and are therefore damped out.
To generate stronger forecast signals, higher quantiles such as 99\% or above,
appear promising (see Fig.~\ref{fig:radarmax} and Tab.~\ref{tab:freqocc}).
This would align with operational forecaster's requests for stronger signals.
Given the current quality of radar data, the proportion of spurious clutter should be well below 1\% of the dataset.

Following the forecaster's feedback, it remains challenging to aggregate ensemble-based probability information so as to meet expectations.

\section*{Acknowledgements}
Thanks to Tam\'{a}s Hirsch, who implemented the processing of the spatial radar quantiles.

\section*{Dataset}
The data that support the findings of this study are openly available at\\
http://doi.org/10.35097/EOvvQEsgILoXpYTK

\bibliographystyle{plain}
\bibliography{references}

\end{document}